\documentclass[aps,prb,groupedaddress,twocolumn,10pt]{revtex4-1}

\begin{document}

\newcommand{\be}{\begin{equation}}
\newcommand{\ee}{\end{equation}}
\newcommand{\bea}{\begin{eqnarray}}
\newcommand{\eea}{\end{eqnarray}}

\title{One SYK SET}

\author{D. V. Khveshchenko}
\affiliation{Department of Physics and Astronomy, University of North Carolina, Chapel Hill, NC 27599}

\begin{abstract}
\noindent
We study the behavior of a single electron transistor (SET) represented by a dissipative 
tunnel junction between a pair of quantum dots described 
by two (possibly, different) Sachdev-Ye-Kitaev (SYK) models. A combined  influence of the soft collective charge and energy  
modes on charge transport is discussed, alongside 
the competing effects of the Coulomb blockade 
and emergent Kondo resonances which might all conspire to result in a non-monotonic behavior of the system's conductance.

\noindent
\end{abstract}
\maketitle

\vspace{0.3cm}
\noindent
{\bf {\it Introduction}} 
\vspace{0.3cm}

The recent transformation [1,2] of the nearly three-decade old SY (Sachdev-Ye) model [3,4] into the novel SYK one has bolstered vigorous activity on a number of important topics. In particular, the asymptotically soluble SYK model is believed to provide an important case study for a genuine holographic (albeit not exactly one-to-one) correspondence and a toy picture of quantum black holes [1,2].  Also, its various higher-dimensional
generalizations, including granular arrays of the coupled SYK systems and their hybrids with other subsystems, can serve as controlled examples of non-Fermi liquids (non-FL) [5]  
and associated phase transitions between such states and more conventional (disordered) Fermi liquids or (many-body) insulators [6]. Besides, the studies of the SYK model rekindled the field of quantum chaos, including the quest for its calculable quantifiers and real-time chaotic dynamics [2]. 

The problem of two interacting SYK systems has also received much attention from the perspective of black hole physics ('traversable wormhole') [7]. As to the more down-to-Earth context of many-body quantum systems, a few recent works [8] addressed the problem of charge 
transport in the SYK 'quantum dot' attached to the ordinary FL leads.  
In the present note we further extend this analysis to the case of a tunnel junction between two (possibly different) SYK systems, one (or both) 
of which might be in the FL state.  

To that end there has been a number of proposals of realizing the SYK physics in the quantum dot/wire  environment [9]. 
In particular, it was argued that the SYK model could serve as a viable description of an irregularly shaped graphene flake in perpendicular magnetic field forcing all the electrons to occupy the highly degenerate lowest Landau level.

The actual SET layout may include additional leads and/or capacitively coupled gates (e.g., single vs two-point contact geometry), as well as multiple modes of transmission.
Nevertheless, its effective theory can often be reduced to that of a single tunnel junction by virtue of projecting out the decoupled  
linear combinations of the fermion modes.

\vspace{0.3cm}
\noindent
{\bf {\it Hamiltonian}} 
\vspace{0.3cm}

A sufficiently general Hamiltonian of such a tunnel junction reads 
\bea 
H=\sum_{\alpha=L,R}
({J^{\alpha}_{i\dots j}\over N_{\alpha}^{(q_{\alpha}-1)/2}}
\psi^{\dagger}_{\alpha i}
\dots
\psi_{\alpha j}
-\mu_{\alpha}\psi^{\dagger}_{\alpha i}
\psi_{\alpha i})
 +
\nonumber\\
\sum_{\alpha,\beta=L,R}
{U_{\alpha\beta}\over 2}
(\psi^{\dagger}_{\alpha i}\psi_{\alpha i}-Q_{\alpha})
(\psi^{\dagger}_{\beta j}\psi_{\beta j}-Q_{\beta})
+
\nonumber\\
{t\over (N_{L}N_{R})^{1/4}}
\sum_n^{N_L}\sum_m^{N_R}(\psi^{\dagger}_{L i}\psi_{R j}
+\psi^{\dagger}_{R j}\psi_{L i})~~~~~~~~~~
\eea
where the complex fermion operators $\psi_{\alpha i}$ correspond to the states localized in the left/right dot and $q_{\alpha}$ are even integers.
The Coulomb charging energy is represented by the (cross)interaction terms $U_{\alpha\beta}$ and also includes the offset charges $Q_{\alpha}$ on the capacitively coupled gates (if any). 

The salient features of the (complex) SYK model can be brought about by treating the amplitudes $J^{\alpha\beta}_{i\dots j}$ of the all-to-all $q_{\alpha}$-fermion couplings as Gaussian random variables with the time- and state-independent variances 
\be 
<J^{\alpha}_{i\dots j}J^{\beta}_{i^{\prime}\dots j^{\prime}}>=\delta^{\alpha\beta}J^2_{\alpha}
\delta_{ii^{\prime}}\dots\delta_{jj^{\prime}} 
\ee
Averaging (1) over such distribution results in introducing time-independent 
$2q$-fermion SYK-type couplings to the action.

In what follows, we consider coupled $SYK_{q}$ models with  (possibly different) indices $q_L$ and $q_R$, one (or both) of which can be in the FL regime formally corresponding to $q=2$.
In that regard, it is worth noting 
that a random $q=2$ term in (1) would result in the same averaged $SYK_2$ coupling 
regardless of whether it represents a random diagonal on-site potential 
($\sum_i\epsilon_i \psi^{\dagger}_i\psi_i$) or an off-diagonal kinetic energy ($\sum_{ij}\epsilon_{ij}\psi^{\dagger}_i\psi_j$).

When Eq.(1) is complemented with the $SYK_2$ $\epsilon$-couplings 
the SYK correlations set in at  $\epsilon^2/J<T\lesssim J$, whereas    
at lower temperatures the system undergoes a crossover to the disordered FL regime [6,10].

Lastly, the $t$-term in (1) describes single-particle tunneling between the dots.
Treating it as fixed is customary in the standard theory of SET. However, one might also consider its amplitude to be random, as in the SYK-lattice models studied in Refs.[6,10].  

The Hamiltonian (1) reveals a number of energy/temperature  scales, the list of which includes
$J_{\alpha}, U_{\alpha\beta}, t^2/J_{\alpha}, J_{\alpha}/N_{\alpha}$, as well as the average single-particle level spacing $\delta_1\sim J/N$
and its many body counterpart $\delta_{N}\sim Je^{-O(N)}$ [1,2].

In the previous work on the conventional SET [11] it was found that 
the onset of the Coulomb blockade (CB) would in general compete with the formation of a single-level (for $T\lesssim\delta_1$) or multi-level (for $T\lesssim\delta_{N}$) Kondo resonances (KR) with the characteristic Kondo temperatures 
 $T_{K1}\sim J e^{-U/\Gamma}$ and $T_{KN}\sim J e^{-U/N\Gamma}$, respectively 
(for $\Gamma<U<J$ where $\Gamma\sim t^2/J$ is the width of the KR). 
 
It was also predicted that the two phenomena (CB and KR) can develop in either order depending on the system's parameters, thus giving rise to a variety of the potential scenarios.
Among them is the 'direct' one where CB would be followed by KR upon lowering the temperature. By contrast, in the 'inverse' scenario the multi-level Kondo scale 
$T_{KN}$ could exceed the Coulomb one and the onset of the Kondo effect can then precede that of the CB. Also, a two-stage Kondo might occur when the Coulomb scale falls right within the range of temperatures between $T_{K1}$ and $T_{KN}$. 

\vspace{0.3cm}
\noindent
{\bf {\it Tunneling as perturbation}} 
\vspace{0.3cm}

From (1) one can readily obtain equations for the fermion propagators
\\
 $G_{\alpha\beta}=\sum_i^{N_{\alpha}}\sum_j^{N_{\beta}}
<\psi_{\alpha i}\psi^{\dagger}_{\beta j}>/{\sqrt {N_{\alpha}N_{\beta}}}$
obeying the equations 
\bea 
(\partial_{\tau}-\mu_{L/R}-\Sigma_{LL/RR}*)G_{LL/RR}-t*G_{RL/LR}=\delta(\tau)\nonumber\\
(\partial_{\tau}-\Sigma_{LL/RR}*)G_{LR/RL}-t*G_{RR/LL}=0~~~~~~~
\eea
where the familiar SYK self-energies [1,2] read
\be
\Sigma_{LL/RR}={J^2_{L/R}}G^{q_{L/R}-1}_{LL/RR}
\ee
and $G_{LR,RL}$ stand for the 
'anomalous' correlators. Notably, 
for $q_L=q_R$ a development of such an off-diagonal order parameter
breaks the $Z_2$ symmetry of the corresponding 'SYK-duplex' model,
as in the wormhole scenario of Refs.[7].

In the absence of the (random) inter-dot SYK correlations ($<J^LJ^R>=0$, as per Eq.(2)) the 'anomalous' propagators can be readily found in the form 
\be
G_{LR/RL}\approx G_{LL/RR}*t*G_{RR/LL}
\ee
and the first equation in (3) for the diagonal propagator $G_{LL/RR}=G_{L/R}$ takes a closed form
\be
G_{L/R}=(\partial_{\tau}-\mu-\Sigma_{L/R})^{-1}
\ee
in terms of the effectively diagonal self-energy 
\be
\Sigma_{L/R}={J^2_{L/R}}G^{q_{L/R}-1}_{L/R}+t*G_{R/L}*t
\ee
The diagonal propagators can then be viewed as a saddle point of the effective theory whose action includes the double time integral  
\be
S_{tun}= {t^2\over 2}\sum_{il}^{N_R}
\sum_{jk}^{N_L}
\int_{\tau_1,\tau_2}\psi^{\dagger}_{Ri}(\tau_1)\psi_{Lj}(\tau_1)\psi^{\dagger}_{Lk}(\tau_2)\psi_{Rl}(\tau_2)
\ee
which replaces the original tunneling term in (1).
This approximation can be further improved, thereby systematically recovering all the (even) higher-order processes, the next one in line ($\sim t^4$) being that of (inelastic) co-tunneling.
 
\vspace{0.3cm}
\noindent
{\bf {\it Conductance: direct tunneling}} 
\vspace{0.3cm}

The charge transport properties of the junction can be  
assessed by computing the current
\be
I=\pm {dQ_{L,R}\over d\tau}=
it\sum_i^{N_L}\sum_j^{N_R}(\psi^{\dagger}_{L i}\psi_{R j}-\psi^{\dagger}_{R j}\psi_{L i})
\ee
whose ensemble average yields the conductivity
$g(T)={dI/dV}|_{V\to 0}$ due to the first-order (direct tunneling) processes  
\be
g_2(T)=
N_LN_R t^2\int d\omega 
{\partial n(\omega)\over \partial\omega}
Im G_L(\omega) Im G_R(-\omega)~~~~~~~
\ee
In what follows, we first neglect the charging energy in (1) and treat the tunneling
term perturbatively while focusing on the mean-field effects of the entangling SYK interactions. 
Such analysis holds for temperatures satisfying the conditions
$U_{\alpha\beta}, t^2/J_{\alpha}, J_{\alpha}/N_{\alpha} < T < J_{\alpha}$ where 
the zero-temperature mean-field fermion propagators read 
\be
G_{\alpha}(\tau\pm 0)=\pm {A_{\alpha}\over \tau^{2\Delta_{\alpha}}}e^{\pm\pi{\cal E}_{\alpha}}
\ee
Here $\Delta_{\alpha}=1/q_{\alpha}$ while the dimensionless parameters ${\cal E}_{\alpha}$ control the fermion occupation numbers in the dots, and the prefactor  
$A_{\alpha}$ is a known function of $q_{\alpha}$ and 
${\cal E}_{\alpha}$ [1,2]. 

A finite-temperature counterpart of (11) can be obtained by the standard conformal transformation 
$\tau\to\sin(\pi\tau T)/\pi T$, although 
Eq.(11) would still suffice for estimating the exponents in any (approximate) 
power-law dependencies.
Computing (10) for $J_{\alpha}/N_{\alpha} < T < J_{\alpha}$
one finds  the linear conductance to the lowest (second) order in $t$
\be
g_2(T)\sim g_0 ({T\over {\sqrt {J_LJ_R}}})^{\eta_{q_L,q_R}}
\ee
where $g_0=\pi N_LN_R t^2/J_{L}J_R$ and the exponent
\be
\eta_{q_L,q_R}=2(\Delta_L+\Delta_R)-2
\ee 
stems from the product of the (vanishing at $\omega\to 0$, except for $q_{\alpha}=2$) densities of states in the coupled dots. 

In particular, if both quantum dots are in the FL regime the conductance is constant ($\eta_{2,2}=0$). In the case of a junction between a FL and $SYK_4$ the exponent in the power-law 
$T$-dependence is $\eta_{2,4}=-1/2$, such value being in agreement 
with the results of Refs.[8]. However, a junction between two $SYK_4$ models would feature the exponent $\eta_{4,4}=-1$. 

\vspace{0.3cm}
\noindent
{\bf {\it SYK fluctuations}} 
\vspace{0.3cm}

As regards the effect of the SYK fluctuations about the mean-field solution (11),
in the no-tunneling/zero-Coulomb limit ($t, U_{\alpha\beta}\to 0$) the Hamiltonian (1) possesses 
a whole manifold of nearly degenerate states which are continuously connected to (11) by virtue of the arbitrary diffeomorphisms of the thermodynamic time 
variable $\tau\to f_{\alpha}(\tau)$,   
obeying the boundary conditions $f_{\alpha}(\tau+\beta)=f_{\alpha}(\tau)+\beta$, combined with the $U(1)$ phase rotations
\be
G_{\alpha}(\tau_1,\tau_2)=A_{\alpha}e^{i\Phi_{\alpha}(\tau_1)-i\Phi_{\alpha}(\tau_2)}
{\large (}
{\partial_{\tau}f_{\alpha}(\tau_1)\partial_{\tau}f_{\alpha}(\tau_2)\over (f_{\alpha}(\tau_1)-f_{\alpha}(\tau_2))^2}{\large)}^{\Delta_q}
\ee 
In addition to being spontaneously broken by a particular choice of the mean-field solution (11) down to the subgroup formed by the Mobius transformations  $SL(2,R)$, the reparametrization 
symmetry is also explicitly violated by the temporal gradients as well as the tunneling and Coulomb terms present in Eq.(1).

At finite temperatures the dynamics of an SYK reparametrization mode is governed by the time integral of the so-called Schwarzian derivative $Sch{\{}\tan{\pi f\over \beta},\tau {\}}$ 
defined according to the formula
$
Sch{\{}y,x {\}}= {y^{\prime\prime\prime}\over y^{\prime}}-{3\over 2}
({y^{\prime\prime}\over y^{\prime}})^2
$
and obeying the differential
'chain rule'
\\
$
Sch {\{}F(y),x{\}}=Sch {\{}F(y),y{\}}{y^\prime}^2+Sch {\{}y,x{\}}
$ [1,2].
  
Upon  the change of variables 
$\partial_{\tau}f_{\alpha}=e^{\phi_{\alpha}}$ this part of the overall action 
takes the form 
\bea
S_{syk}=\sum_{\alpha}{\gamma_{\alpha}N_{\alpha}\over J_{\alpha}}
\int_{\tau}({1\over 2}({\partial_{\tau}\phi_{\alpha}})^2-({2\pi\over \beta})^2e^{2\phi_{\alpha}})
\eea
where the $q_{\alpha}$-dependent prefactor $\gamma_{\alpha}$ was computed numerically (see Refs.[2] and references therein). In the FL case of $q_{\alpha}=2$ one has $\gamma_{\alpha}=0$.  

Thus, the fluctuations of the soft $\phi_{\alpha}$ modes develop at energies below $J_{\alpha}/N_{\alpha}$, so at higher temperatures their effect can be neglected. 
In contrast, for $T<J_{\alpha}/N_{\alpha}$ the 'gravitational dressing' 
of any product of the vertex operators $e^{\phi_{\alpha}(\tau)}$ can be performed 
in the basis of the  eigenstates of the underlying quantum mechanical hamiltonian of the 
Liouville  theory deformed with the 'quench' potential acting during 
the time intervals between consecutive insertions of such operators [10,12].  
As the result, an arbitrary power $p$ of the two-point propagator 
of an isolated SYK system develops a universal asymptotic behavior
\be
<G^p_{\alpha}(\tau)>\sim 1/(J_{
\alpha}\tau)^{3/2}  
\ee
for all positive integer $p$ and $q_{\alpha}>2$  [10,12].

Applying this result to Eq.(12) one finds that as long as the CB effects remain
negligible the lowest-order conductivity develops a linear temperature dependence ($\eta_{q_L,q_R}=1$) for all $q_{\alpha}>2$ in the entire range $E_c < T < J_{\alpha}/N_{\alpha}$).
However, for $q_L=2$ the corresponding exponent is $\eta_{2,4}=1/2$. 

It is worth mentioning, though, that an apparent agreement between the above exponent and the value obtained for $q_L=2, q_R=q$ in the first of Refs.[8] for the 
non-linear conductance $g(V)$ in the range $t^2/J, J/N < V < J$ is no more than an accident.
In fact, the latter  pertains to the limit of strong tunneling where the propagators $G_L/R$ in (10) would have to be replaced with 
$
G_{L/R}/(1\pm ig_0G_{R/L})
$.

Consequently, the conductance would turn out to be proportional to $1/g_0$, thereby inverting the Fourier transform of the long-$\tau$ asymptotic (11), while accounting for neither the SYK fluctuations, nor the effects of CB.
    
Nonetheless, the conductance $g(V)$ computed in [12]
demonstrates an interesting duality, changing from $\eta_{2,q}=2/q-1$ at weak tunneling,  
$(t^{2q}/J^{4})^{1/(2q-4)}<V$, 
\\
 to $\eta_{2,q}=1-2/q$ in the complementary strong tunneling regime. 

Lastly, in the general case of a large disparity between the (both non-vanishing) $\gamma_LJ_L/N_L$ and $\gamma_RJ_R/N_R$ the exponent (13) approaches $\eta_{q_{L},q_{R}}=2/q_{L/R}-1/2$ for
$J_{L/R}/N_{L/R} < T < J_{R/L}/N_{R/L}$ since the SYK fluctuations affect only one of the two dots. 

Moreover, with increasing strength of the coupling between the dots an independent application of two different $Diff(S^1)$ transformations would conflict with the tunneling term. Such a strong symmetry-breaking effect could be minimized, though, if the $L/R$ reparametrizations were locked 
into one common transformation $f_L=f_R$, thereby resulting in the universal behavior of the averaged products of an arbitrary number of $G_L$ and $G_R$ alike, $<G_L^{p_L}G_R^{p_R}>\sim 1/\tau^{3/2}$. 

\vspace{0.3cm}
\noindent
{\bf {\it Charging effects}} 
\vspace{0.3cm}

In addition to the Schwarzian modes, for a sizable CB energy $E_c>J_{\alpha}/N_{\alpha}$ 
the mean-field results become strongly affected by the charge fluctuations. According to the earlier studies of
the conventional SET and other quantum devices 
comprised of the total of up to four dots and/or leads [11,13]
the analysis of such fluctuations can be facilitated by introducing the standard representation of the fermion operator as a product of its energy and charge degrees of freedom
$
\psi_{\alpha n}=\chi_{\alpha n}e^{i\Phi_\alpha}
$.
In this way, the fermionic particle-hole excitations in the dots become separated from the collective ('plasmon') degrees of freedom.

While the 'fractionalized' fermionic degrees of freedom $\chi_{\alpha n}$ can still be traded for the bi-local
collective variables $G_{\alpha}$ and $\Sigma_{\alpha}$ those are now coupled to the phase fields 
$\Phi_{\alpha}$.   
The relevant part of the action then takes the form $S=S_{syk}+S_{\Phi}$ where the two terms describe (approximately) uncoupled fluctuations of the fields $\phi_{\alpha}$ and $\Phi_{\alpha}$, respectively.

In particular, in the regime dominated by CB, $J_{\alpha}/N_{\alpha}<T<E_c$, the SYK fluctuations remain frozen,  
while the dynamics of $\Phi_{\alpha}$ is described in terms of     
the (Euclidean) 'phase-only' effective action for the antisymmetric (out-of-phase) combination 
$\Phi=\Phi_{L}-\Phi_{R}$ (while the symmetric one decouples from the fields $\phi_{\alpha}$)
and the concomitant parameters $\Delta Q$, ${\Delta\cal E}$ 
\bea
S_{\Phi}={1\over 2E_c}\int_{\tau}({\partial_{\tau}\Phi}-i{E_c\Delta Q\over e}+2\pi i T\Delta
{\cal E})^2\nonumber\\
+
{1\over 2}\int_{\tau_1}\int_{\tau_2} K(\tau_1-\tau_2)\cos(\Phi(\tau_1)-\Phi(\tau_2))
\eea
The effective Coulomb energy $E_c$ includes a contribution  
stemming from the time derivatives in the action of the complex SYK model which explicitly breaks the $U(1)$ symmetry (alongside the $Diff(S^1)$ one) and is proportional to 
the inverse compressibility $\kappa$ of the multi-fermion SYK system, $E_c=U+\kappa^{-1}$ [14].

As previously mentioned, the last term in Eq.(17) is the leading one in  
the systematic expansion of the effective bosonic action in powers of $t^2$. 
Alternatively, in the case of a random inter-dot hopping amplitude $t$ it results from averaging over the Gaussian distribution, $<t^2>=t^2$. 
 
The corresponding kernel 
\be
K(\tau)=t^2 G_L(\tau) G_R(-\tau)D(-\tau)
\ee
represents the dissipative medium of the SYK particle-hole excitations. It  
is to be determined self-consistently in terms of the 
phase field propagator 
\\
$D(\tau)=<e^{i\Phi(\tau)}e^{-i\Phi(0)}>$ given by the formula
\be 
D(\tau)=(\partial_{\tau}^2/E_c+\lambda+K)^{-1}
\ee
where the 
Lagrange multiplier $\lambda$ enforces the normalization condition $D(0)=1$.

It is worth mentioning that in the regime of interest 
a strongly non-Gaussian nature of the action (17) prevents one from computing the correlator $D(\tau)$ simply as the Debye-Waller factor $\exp(-<\Phi(\tau)\Phi(0)>/2)$. 

Also, the functional integral with the action (17) is to be evaluated on the field distributions 
subject to the boundary condition $\Phi(\tau+\beta)=\Phi(\tau)+2\pi n$
which reflects the compactness of the phase variable $\Phi$. Therefore, for large phase fluctuations it is essential to account for the effect of the topological  
'$\theta$-term' [15] 
\be
S_{top}=-i(\Delta {\tilde Q}/e)\int_{\tau}\partial_{\tau}\Phi
\ee
embedded in (17) which is purely imaginary and 
proportional to the difference $\Delta{\tilde Q}=\Delta Q-2\pi eT\Delta{\cal E}/E_c $ between the offset charges on the gates capacitively coupled to the dots (if any),  
corrected for the $T$-dependent excess charges on the dots themselves (notably, $2\pi{\cal E}_{\alpha}=-\partial\mu_{\alpha}/\partial T|_{T\to 0}$ [1,14]). 

\vspace{0.3cm}
\noindent
{\bf {\it Phase dynamics}} 
\vspace{0.3cm}

Importantly, in the regime where the kernel demonstrates an algebraic behavior 
\be
K(\tau)=g_0/\tau^{1+s}
\ee
governed by the exponent
\be
s=\Delta_L+\Delta_R
\ee
it appears to be generically sub-Ohmic ($s<1$),
as long as, at least, one of $q_{L/R}$ is greater than $2$.

This should be contrasted against the ordinary (FL) SET 
where attaining such a regime would only be possible in the presence of sufficiently strong excitonic enhancement in the final state. Otherwise, the behavior of the kernel (21) would turn out to be super-Ohmic due to the competing effect of orthogonality catastrophe [13]). 

The earlier analyses of the sub-Ohmic phase-only model with the kernel (21) 
demonstrated that at the critical coupling the conductivity takes its maximal (temperature independent) value which was estimated in Refs.[13] as 
\be
g_{c}\approx {1\over 2\pi^2(1-s)}
\ee
At this critical point the system undergoes a second-order quantum phase transition from the disordered ($<\cos\Phi>=0$) 
CB-governed phase for $g<g_c$
to a dissipation-driven ordered ($<\cos\Phi>\neq 0$) conducting one for $g>g_c$.  

In the critical regime the system of coupled equations (6), (7) where the term $t*G_{R/L}*t$ now contains the extra factor $D(\tau)D(-\tau)$, and (19)  
permits a scaling-invariant solution.
Namely, $G_{L/R}$ retain 
their SYK behavior (11)  
while dragging along the phase correlator featuring a power-law asymptotic $D(\tau)=B/\tau^{2\Delta_{\Phi}}$ with the exponent
\be
\Delta_{\Phi}={1\over 2}(1-s)
\ee
and the prefactor $B\sim(t^2A_LA_R)^{-1}$.
Correspondingly, the amplitudes $A_{L,R}$ get reduced, as per the equation
$
J^2A_{\alpha}^{q_{\alpha}}=1-t^2BA_{L}A_R
$. 

In the ordered current-carrying resistive regime developing for $g>g_{c}$ the phase field gets condensed and its propagator reaches a finite limit $D(\tau\to\infty)=const$ corresponding to the vanishing effective charging energy $E_c^{*}$, consistent with the slower than $1/\tau^2$ power-law decay of the kernel (21). 

By contrast, in the insulating phase the field $\Phi$ 
is disordered (while the charge  dual to $\Phi$ tends to become quantized, $\Delta Q=n$), its fluctuations are gapped, and their correlator decays exponentially 
\be
D(\tau)=e^{-E^{*}_c/2T}F(T\tau, {\cal E})
\ee
where the factor 
$F=\sum_ne^{-E_cn^2/2+2\pi n{\cal E}-nT\tau}$
\\
stems from the infinite sum over the arbitrary winding numbers. 

The renormalized Coulomb gap can then be estimated from 
the normalization condition $D(0)=1$ 
which yields 
\be
E_c^{*}=<\lambda>=E_c(1-g/g_{c})^{\nu}
\ee
where $\nu=1/(1-s)$ for $s>1/2$ and  $\nu=1/s$ for $s<1/2$ [13].
For comparison, in the marginal case $s=1$ the renormalized charging energy would be reduced, yet remain finite
($
E_c^{*}=2\pi g^2 E_ce^{-\pi g}
$
and 
$
E_c^{*}=E_c(1-4g/\pi+\dots)
$
for $g>>1$ and $g<<1$, respectively [13]).

Also, the behavior of $E_c^{*}$ (hence the conductance) strongly depends on the offset gate charges. 
In the previous works [8] this dependence was not addressed, as if $\Delta Q$ assumed the default value of zero. 

\vspace{0.3cm}
\noindent
{\bf {\it Conductance: inelastic co-tunneling}} 
\vspace{0.3cm}

At the transition points between the charge quantization plateaus ($Q/e=n+1/2$) the bare gap vanishes and the conductance computed in the 'phase-only' theory (17) appears to increase indefinitely upon lowering temperature. However, by  invoking the Friedel sum rule one might argue that the conductance attains a finite value given by the total number of transmission channels (which may depend on the setup's configuration) and corresponding to the unitarity limit [11,13].

In contrast, on the charge quantization plateaus $\Delta Q=n$ the renormalized Coulomb energy is maximal, so in the CB regime  $J_{\alpha}/N_{\alpha}<T<E_c$ 
the averaged current acquires the extra $D^2(\tau)$ factor. Accordingly,
the conductance (12) gets suppressed by an additional activation-like factor, 
$g_2(T)\sim g_0e^{-E^{*}_c/T}(T/J)^{\eta}$.

The leading temperature dependence will then be dictated by the next (forth) order contribution 
corresponding to the co-tunneling processes which contribute to the current as
\be
g_4(T)=N_LN_R {t^4}
\int d\omega {\partial n(\omega)\over \partial\omega}\\
Im[KD]_{\omega}
Im[KD]_{-\omega}
\ee
Converting the integral (27) into the time domain one can see that it escapes the activation-like suppression due to the integration over two short ($\sim 1/E_c^*$, as opposed to the typically longer, $\sim 1/T$) times. It then dominates the power-law behavior of the conductivity 
\be
g_4(T)\sim g^2_0 {J_LJ_R\over E_c^2}({T\over {\sqrt {J_LJ_R}}})^{\eta^{\prime}_{q_L,q_R}}
\ee
with the exponent 
\be
\eta^{\prime}_{q_L,q_R}={4(\Delta_L+\Delta_R)-2}
\ee
In the case of $q_L=2, q_R=4$, one obtains the linear $T$ dependence ($\eta^{\prime}_{2,4}=1$) applicable for $J_R/N_R<T<E_c$ (the parameters $J_L, N_L$ of the dot in the FL state do not affect the range of applicability) which then dominates over the CB-suppressed direct tunneling (12). 

At still lower temperatures, $T<J_R/N_R$, this behavior will crossover to another asymptotic with $\eta^{\prime}_{2,4}=3/2$. Both of the above regimes were predicted in the third of Refs.[8]. Either power-law 
would then be markedly different from that in the ordinary (FL) quantum dots 
with the standard $\eta^{\prime}_{2,2}=2$ behavior, independent of $J_{\alpha}$. 

It is worth mentioning, though, that while the complementary elastic counterpart of the co-tunneling conductance is sensitive to the details of the actual fermion motion through the dots (hence, the phase of its wave function) it may result in a $T$-independent contribution, akin to the FL case where the conductance saturates at $g_{4}\sim g^2_0\delta_1/E_c$ [16].

Moreover, at low temperatures ($T<J_{\alpha}/N_{\alpha}$) the 'gravitational dressing' with the $\phi_{\alpha}$ fluctuations responsible for the above hallmarks of the SYK physics may eventually be forestalled by the developing KR [11].  
  
In the FL case, the single-level KRs start to form upon lowering the temperature past the average single-particle level spacing $\delta_1$. Incidentally,
the Schwarzian fluctuations begin to develop at the same $T\sim\delta_1$ (possibly, up to a factor $\sim 1/\ln N$ [2,12]). Regardless of whether or not the single-level FL scenario of the KR remains intact in the SYK systems, it can still develop below the characteristic multi-level Kondo scale $T_{KN}$ where coherent transport through the SET can be supported by a many-body resonance close to the Fermi level. In this regime the conductance, too, tends to the unitarity limit as $T\to 0$.  

Given such an interplay of the competing CB, KR, topological term (20), and sub-Ohmic 'dissipation' manifested by the kernel (17), the conductance $g(T)$ would generally be non-monotonic and showing several 
regimes, ranging from an activation type of behavior to an algebraic decay (or increase, depending).

\vspace{0.3cm}
\noindent
{\bf {\it Discussion}} 
\vspace{0.3cm}

The overall picture of charge transport in the coupled SYK dots can then be described as follows. 
At high temperatures ($T>J, E_c$) the entire spectrum of an isolated SYK quantum dot appears 
nearly $N$-fold degenerate with an effective width of order $J$. 
The fermion spectral weight can then be approximated by a broadened peak, thereby resulting to the behavior akin to that in the multi-level Kondo effect. The conductance gradually increases with decreasing $T$. 

However, once the temperature drops below $J$ but remains above $E_c$ the SYK physics starts to set in. For as long as the phase $\Phi$ fluctuations remain suppressed it manifests 
itself in the power-law conductance (12) which continues to rise.
   
At lower temperatures, $J/N\lesssim T\lesssim E_c$,  
the sub-Ohmic phase dynamics gets activated and the behavior of the conductance starts to non-trivially depend on the coupling strength $g_0$. However, the reparametrization mode remains frozen and the mean-field SYK description can still be used. 

In particular, at $g_0<g_{c}$ with $g_{c}$ given by Eq.(23) the system is in the CB regime with the exponentially decaying $g_2(T)$. At $g_0>g_{c}$, however, 
$E^{*}_c$ gets renormalized all the way down to zero, the phase variable remains condensed, and the conductivity continues to increase while the dependence on $Q$ can only be observed at some intermediate temperatures.

At still lower temperatures, $T\lesssim J/N$, the fluctuations of the reparametrization mode become fully developed, thereby renormalizing the arbitrary products of $G_{L/R}$ in the universal manner. Regardless of whether it happens on just one or both sides of the junction the effective kernel (17) becomes super-Ohmic ($s>1$) and then CB ensues for all $g_0$,  except for the transitions between the plateaus at $Q=n+1/2$ where $E^{*}_c$ vanishes. 

Eventually, the CB regime gets arrested below $T_K$ where the formation of KRs takes over, the conductivity once again reversing its behavior.  
It is possible, though, that such a recovery gets delayed until much lower temperatures set by the many-body level spacing $\delta_{N}$, rather than its FL single-level counterpart $\delta_1$, due to the presence of the strong SYK correlations.

A further analysis of this convoluted interplay between the SYK, charging, Kondo, and tunneling effects will be presented elsewhere. It would also be of interest to extend it to the full counting statistics and its properties, as in the first of Refs.[8], as well as the other physically relevant observables.

\end{document}